\documentstyle[twocolumn,aps,prl,epsf,psfig]{revtex}
\begin{document}
\def\r{{\bf{r}}}
\def\i{{\bf{i}}}
\def\j{{\bf{j}}}
\def\m{{\bf{m}}}
\def\k{{\bf{k}}}
\def\kt{\tilde{\k}}
\def\K{{\bf{K}}}
\def\P{{\bf{P}}}
\def\q{{\bf{q}}}
\def\Q{{\bf{Q}}}
\def\p{{\bf{p}}}
\def\x{{\bf{x}}}
\def\X{{\bf{X}}}
\def\Y{{\bf{Y}}}
\def\F{{\bf{F}}}
\def\G{{\bf{G}}}
\def\M{{\bf{M}}}
\def\V{\cal V}
\def\tchi{\tilde{\chi}}
\def\tk{\tilde{\bf{k}}}
\def\tK{\tilde{\bf{K}}}
\def\tq{\tilde{\bf{q}}}
\def\tQ{\tilde{\bf{Q}}}
\def\si{\sigma}
\def\ep{\epsilon}
\def\al{\alpha}
\def\be{\beta}
\def\ep{\epsilon}
\def\up{\uparrow}
\def\de{\delta}
\def\De{\Delta}
\def\up{\uparrow}
\def\dwn{\downarrow}
\def\ksi{\xi}
\def\etha{\eta}
\def\product{\prod}
\def\goto{\rightarrow}
\def\switch{\leftrightarrow}

\title{Pseudogaps in the 2D half-filled Hubbard model}
\author{C.\ Huscroft, M. Jarrell, Th.\ Maier, S.\ Moukouri, A.N.\ Tahvildarzadeh}
\address{Department of Physics, University of Cincinnati, 
Cincinnati, OH 45221-0011}
\date{\today }
\maketitle

\begin{abstract}
        We study the pseudogaps in the spectra of the half-filled 2D Hubbard 
model using both finite-size and dynamical cluster approximation (DCA) quantum 
Monte Carlo calculations.  A charge pseudogap, accompanied by non-Fermi liquid 
behavior in the self energy, is shown to persist in the thermodynamic limit.  
The DCA (finite-size) method systematically underestimates (overestimates) 
the width of the pseudogap.  A spin pseudogap is not seen at half-filling.  

\end{abstract}


\paragraph*{Introduction}  For over a decade it has been recognized
that the normal state properties of high-$T_c$ superconductors
are unusual and appear to have non-Fermi liquid characteristics.\cite{maple1}  
One of the most remarkable features of the normal state is a suppression
of the density of states at the Fermi energy in a temperature
regime above $T_c$ in underdoped samples.  Angular resolved photoemission 
experiments\cite{ding,ronning} show that this pseudogap in the
spectral function has a d-wave anisotropy, the same symmetry as
the superconducting order parameter in these materials.  This, along
with theories that short-ranged spin fluctuations mediate pairing 
in the high-$T_c$ cuprates\cite{scalapino1,timusk}, emphasizes the
importance of understanding the normal state, insulating phase.

It is thought by many that the two-dimensional Hubbard model, or closely
related models, should capture the essential physics of the high-$T_c$
cuprates.~\cite{scalapino1}  Yet, despite years of effort,
neither the precursor pseudogap nor d-wave 
superconducting order have been conclusively seen in the Hubbard model.

Intuitively, one may expect that the Hubbard model should show
pseudogap behavior.  At half-filling, the ground state of the 2D
Hubbard model is an antiferromagnetic insulator\cite{hirsch1,white1}
and the spectrum is therefore gapped.  However, the Mermin-Wagner 
theorem precludes any transition at finite $T$, so as the temperature
is lowered one may anticipate that a pseudogap will develop.\cite{kampf1}
This question has been previously addressed in the 2D Hubbard insulator 
by finite-size lattice Quantum Monte Carlo (QMC)\cite{white2,creffield} 
and approximate many-body techniques \cite{deisz,vilk,moukouri}.  The 
results have been contradictory and inconclusive as to the existence of 
a pseudogap at low temperatures, due to limitations of these techniques.

Using the recently developed Dynamical Cluster Approximation 
(DCA)\cite{DCA_hettler,DCA_maier1} we 
find that at sufficiently low temperatures a pseudogap opens in the 
single particle spectral weight $A({\bf k},\omega)$ of the 2D Hubbard 
model with a simultaneous destruction of the Fermi liquid state due to 
critical fluctuations above the $T=0$ transition temperature.  This 
occurs in the weak-to-intermediate coupling regime $U<W$, where $U$ is 
the on-site Coulomb energy and $W$ the non-interacting band width. 

Using finite-sized techniques, it is difficult to determine if a gap
persists in the thermodynamic limit.  At half filling, finite-size QMC 
calculations display a gap in their spectra as soon as the correlation 
length exceeds the lattice size, so they tend to overestimate the pseudogap
as it would appear in the thermodynamic limit.  
Finite-size scaling is complicated by the lack of an exact scaling ansatz
for the gap and the cost of performing simulations of large systems.
Calculations employing Dynamical Mean Field Approximation 
(DMFA)\cite{metzvoll} in the paramagnetic phase do 
not display this behavior since they take place in the thermodynamic limit
rather than on a finite-size lattice.  However, the DMFA lacks the non-local 
spin fluctuations often believed to be responsible for the pseudogap.  
The Dynamical Cluster Approximation (DCA) is a fully causal approach which 
systematically incorporates non-local corrections to the DMFA by mapping 
the problem onto an embedded impurity cluster of size $N_c$.  $N_c$ 
determines the order of the approximation and provides a systematic 
expansion parameter $1/N_c$.  While the DCA becomes exact in the limit of 
large $N_c$ it reduces to the DMFA for $N_c=1$.  Thus, the DCA differs from 
the usual finite size lattice calculations in that it is a reasonable 
approximation to the lattice problem even for a ``cluster'' of a single 
site.  Like the DMFA, the DCA solution remains in the thermodynamic limit, 
but the dynamical correlation length is restricted to the size of the 
embedded cluster.  Thus the DCA tends to underestimate the pseudogap.

\begin{figure}[ht]
\leavevmode\centering\psfig{file=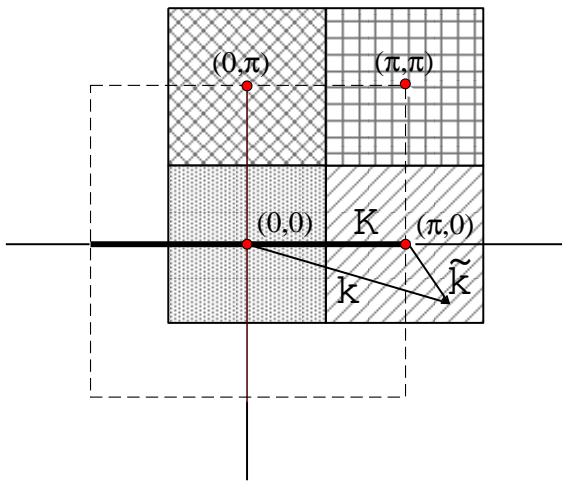,width=3. in}
\caption{$N_c=4$ cluster cells (shown by different fill patterns) that 
partition the first Brillouin Zone (dashed line).  Each cell is centered
on a cluster momentum $\K$ (filled circles). To construct the DCA cluster, we 
map a generic momentum in the zone such as $\k$ to the nearest cluster point 
$\K=\M(\k)$ so that $\kt=\k-\K$ remains in the cell around $\K$.  }
\label{BZ}       
\end{figure}

\paragraph*{Method}
The DCA is based on the assumption that the lattice self energy is weakly
momentum dependent.  This is equivalent to assuming that the dynamical 
intersite correlations have a short spatial range $b \alt L/2$ where $L$ is 
the linear dimension of the cluster.  Then, according to Nyquist's sampling 
theorem\cite{nyquist}, to reproduce these correlations in the self energy, 
we only need to sample the reciprocal space at intervals of 
$\Delta k\approx 2\pi/L$.  Therefore, we could approximate $G(\K+\kt)$ by 
$G(\K)$ within the cell of size $\left( \pi/L\right)^D$ (see, Fig.~\ref{BZ}) 
centered on the cluster momentum $\K$ (wherever feasible, we suppress the 
frequency labels) and use this Green function to calculate the self energy.  
Knowledge of these Green functions on a finer scale in momentum is unnecessary, 
and may be discarded to reduce 
the complexity of the problem. Thus the cluster self energy can be constructed
from the {\em{coarse-grained average}} of the single-particle Green function 
within the cell centered on the cluster momenta:
\begin{equation}
\bar{G}(\K) \equiv \frac{N_c}{N}\sum_{\kt}G(\K+\kt),
\label{gbar}
\end{equation}
where $N$ is the number of points of the lattice, $N_c$ is the number of 
cells in the cluster, and the $\kt$ summation runs over the momenta of 
the cell about the cluster momentum  $\K$ (see, Fig.~\ref{BZ}).  For short 
distances $r\alt L/2$ the Fourier transform of the Green function 
$\bar{G}(r) \approx G(r) +{\cal{O}}((r\Delta k)^2)$, so that short ranged 
correlations are reflected in the irreducible quantities constructed from 
$\bar{G}$; whereas, longer ranged correlations $r>L/2$ are cut off by the 
finite size of the cluster.\cite{DCA_hettler}

This coarse graining procedure and the relationship of the DCA to the DMFA 
is illustrated by a microscopic diagrammatic derivation of the DCA.  For 
Hubbard-like models, the properties of the bare vertex are completely 
characterized by the Laue function $\Delta$ which expresses the momentum 
conservation at each vertex.  In a conventional diagrammatic approach
$\Delta(\k_1,\k_2,\k_3,\k_4)= 
\sum_\r \exp\left[i\r\cdot(\k_1-\k_2+\k_3-\k_4)\right]=
N \delta_{\k_1+\k_2,\k_3+\k_4}$ where $\k_1$ and $\k_2$ ($\k_3$ and 
$\k_4$) are the momenta entering (leaving) each vertex through its
legs of $G$.  However as $D\to\infty$ M\"uller-Hartmann showed that the  
Laue function reduces to\cite{muller-hartmann}
\begin{eqnarray}
\Delta_{D\rightarrow\infty}({\bf k}_1,{\bf k}_2,{\bf k}_3,{\bf k}_4)=
1+{\cal O}(1/D)\quad\mbox{.}
\label{Laueinft}
\end{eqnarray}
The DMFA assumes the same Laue function, 
$\Delta_{DMFA}(\k_1,\k_2,\k_3,\k_4)=1$, 
even in the context of finite dimensions.  Thus, the conservation of momentum 
at internal vertices is neglected.  
Therefore we may freely sum over the internal momenta at each vertex in the 
generating functional $\Phi_{DMFA}$. This leads to 
a collapse of the momentum dependent contributions to the functional 
$\Phi_{DMFA}$ and only local terms remain.

The DCA systematically restores the momentum
conservation at internal vertices.  As discussed above the Brillouin-zone 
is divided into $N_c=L^D$ cells of size $(2\pi/L)^D$. 
Each cell is represented by a cluster momentum $\bf K$ in the center of 
the cell. We require that momentum conservation is (partially) observed 
for momentum transfers between cells, i.e. for momentum transfers larger 
than $\Delta k=2\pi/L$, but neglected for momentum transfers within a 
cell, i.e less than $\Delta k$. This requirement can be established by 
using the Laue function \cite{DCA_hettler}
\begin{equation}
\Delta_{DCA}(\k_1,\k_2,\k_3,\k_4)=
N_c \delta_{\M(\k_1)+\M(\k_3),\M(\k_2)+\M(\k_4)}
\label{lauedca}
\end{equation}
where $\M(\k)$ is a function which maps $\k$ onto the momentum label $\K$
of the cell containing $\k$ (see, Fig.~\ref{BZ}).    

\begin{figure}[ht]
\leavevmode\centering\psfig{file=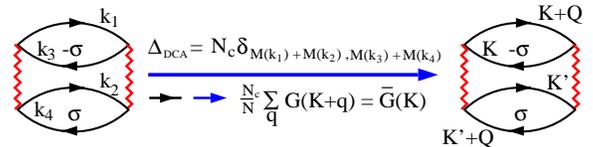,width=3. in}
\caption{  The DCA choice of the Laue function Eq.~\ref{lauedca} leads to
the replacement of the lattice propagators $G({\bf k}_1)$, 
$G({\bf k}_2)$,... by coarse grained propagators $\bar{G}({\bf K})$,
$\bar{G}({\bf K}^\prime)$, ... (Eq.~\ref{gbar}) in the internal 
legs of $\Phi_{DCA}$, illustrated for a second order diagram.
}
\label{collapse_DCA}       
\end{figure}

With this choice of the Laue function the momenta of each internal leg may
be freely summed over the cell.  This is illustrated for the second-order 
term in the generating functional in Fig.~\ref{collapse_DCA}.  Thus, each 
internal leg $G(\k_1)$ in a diagram is replaced by the coarse--grained Green 
function ${\bar G}(\M(\k_1))$, defined by Eq.~\ref{gbar}
\noindent The diagrammatic sequences for the generating functional and its 
derivatives are unchanged; however, the complexity of the problem is 
greatly reduced since $N_c\ll N$.  We showed 
previously\cite{DCA_hettler,DCA_maier1} that the DCA estimate of 
the lattice free-energy is minimized by the approximation 
$\Sigma(\k)\approx {\bar{\Sigma}}({\bf M}({\bf k}))$,
where $\delta\Phi_{DCA}/\delta\bar{G}=\bar{\Sigma}$.

The cluster problem is then solved by usual techniques such as 
QMC\cite{hossein}, the non-crossing approximation\cite{DCA_maier1} or
the Fluctuation-Exchange approximation.  Here we employ a 
generalization of the Hirsh-Fye QMC algorithm\cite{fye} to solve
the cluster problem.  The initial Green function for this procedure
is the bare cluster Green function ${\cal{G}}(\K)^{-1}=\bar{G}(\K)^{-1}+
{\bar{\Sigma}}(\K) $ which must be introduced to avoid over-counting
diagrams.  The QMC estimate of the cluster self
energy is then used to calculate a new estimate of $\bar{G}(\K)$
using Eq.~\ref{gbar}.  The corresponding ${\cal{G}}(\K)$ is used
to reinitialize the procedure which continues until the self energy
converges to the desired accuracy.

\begin{figure}[ht]
\leavevmode\centering\psfig{file=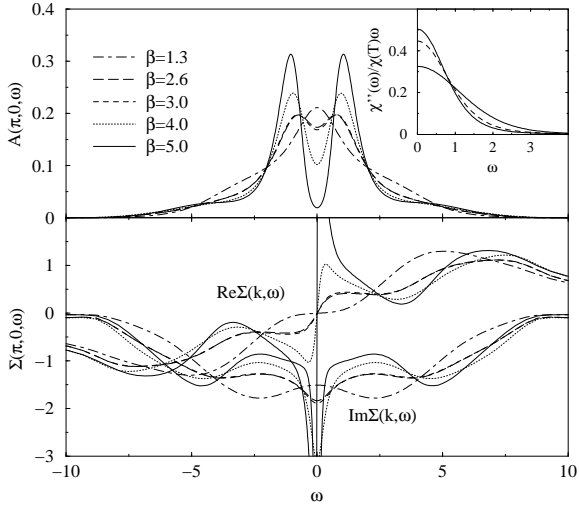,angle=-90,width=3. in}
\caption{The spectral density $A({\bf k}, \omega)$, and the real 
${\rm Re}\Sigma({\bf k},\omega)$ and imaginary 
${\rm Im}\Sigma({\bf k},\omega)$ parts of the self-energy for the 
2D Hubbard model via the DCA with a paramagnetic host
at ${\bf k} = (\pi, 0)$ for a 64-site cluster ($N_c = 64$)
at various temperatures.  The on-site Coulomb repulsion $U=5.2$, the 
band width $W=8$, and the filling $\langle n \rangle = 1$.  As the 
temperature is lowered, the system first builds a Fermi-liquid-like
peak in $A({\bf k}, \omega)$.  By $\beta = 2.6$, a pseudogap begins 
to develop in $A({\bf k}, \omega)$ and simultaneously, 
${\rm Re}\Sigma({\bf k},\omega)$ develops a positive slope at 
$\omega=0$, a signal of a non-Fermi liquid.  The pseudogap deepens as 
the temperature is further lowered.   The imaginary part of the
dynamic spin susceptibility, divided by the static spin susceptibility
is shown in the inset.  No spin gap is seen.
}
\label{pizero}       
\end{figure}

\paragraph*{Results}  We study the 2D Hubbard Hamiltonian:
\begin{eqnarray}
H &=& - t \sum_{\langle i,j \rangle, \sigma}
(c_{i\sigma}^{\dagger} c_{j\sigma}
+ c_{j\sigma}^{\dagger} c_{i\sigma})  \nonumber \\
&&+ U \sum_{i} (n_{i\uparrow} - \frac12)
(n_{i\downarrow} - \frac12) 
-\mu \sum_{i,\sigma} n_{i\sigma} \, .
\label{hubham}
\end{eqnarray}
where $c_{i\sigma}^{\dagger} (c_{i\sigma})$ creates (destroys) an
electron at site $i$ with spin $\sigma$, $U$ is the on-site Coulomb
potential, and $n_{i\sigma} = c_{i\sigma}^{\dagger} c_{i\sigma}$ 
is the number operator.  We set the overlap integral $t=1$ and 
measure all energies in terms of $t$.  We work at $\mu=0$ where the 
system is half-filled ($\langle n \rangle = 1$). We choose $U=5.2$, 
which is well below the value $U\agt W$ believed to be necessary to
open a Mott-Hubbard gap. 

We also calculate the angle integrated dynamical spin susceptibility shown 
in the inset. It does not have a pseudogap, as expected for the half-filled 
model since the spin-wave spectrum is gapless.  Since a (spin) charge gap 
is generally defined as one which appears in the (spin) charge dynamics or 
thermodynamics, we conclude that the pseudogap is only in the charge 
response and is due to short-ranged antiferromagnetic spin correlations.  

Fig.~\ref{pizero} shows the spectral density $A({\bf k}, \omega)$,
and the real ${\rm Re}\Sigma({\bf k},\omega)$ and imaginary 
${\rm Im}\Sigma({\bf k},\omega)$ parts of the self-energy for the 
2D Hubbard model via the DCA with a paramagnetic host
at the Fermi surface $X$ point ${\bf k} = (\pi, 0)$ for a 
64-site cluster ($N_c = 64$) at various temperatures.  
We obtain the spectral function $A({\bf k}, \omega)$ via the
Maximum Entropy Method (MEM).\cite{JARRELLandGUB}
As the temperature is lowered, the system first builds a Fermi-liquid-like
peak in $A({\bf k}, \omega)$.
By $\beta = 2.6$, a pseudogap begins to develop in $A({\bf k}, \omega)$.
The pseudogap builds as the temperature is further lowered.  

\begin{figure}[ht]
\leavevmode\centering\psfig{file=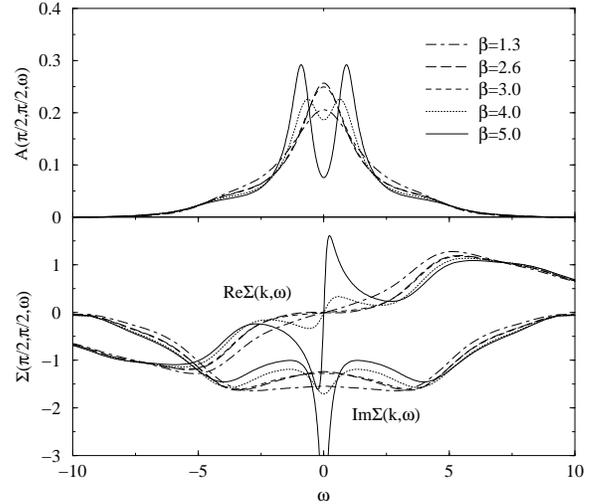,angle=-90,width=3. in}
\caption{The spectral density $A({\bf k}, \omega)$, and the real 
${\rm Re}\Sigma({\bf k},\omega)$ and imaginary 
${\rm Im}\Sigma({\bf k},\omega)$ parts of the self-energy for the 
2D Hubbard model via the DCA at ${\bf k} = (\pi/2, \pi/2)$ and the 
same parameters as Fig.~\ref{pizero}.  Again the system first builds 
a Fermi-liquid-like peak in $A({\bf k}, \omega)$ and then develops a 
pseudogap in $A({\bf k}, \omega)$ with a simultaneous non-Fermi liquid 
behavior in ${\rm Re}\Sigma({\bf k},\omega)$.  Here, though, the 
pseudogap first appears at a lower temperature than at 
${\bf k} = (\pi, 0)$.}
\label{pio2pio2}       
\end{figure}

Fig.~\ref{pio2pio2} shows the spectral function $A({\bf k}, \omega)$ at 
the half-filled Fermi surface point ${\bf k} = (\pi/2, \pi/2)$.  The 
qualitative features are similar to Fig.~\ref{pizero}, but the pseudogap 
opens at a lower temperature and the distance between the peaks is less 
than that seen at the $X$ point.  This behavior is reminiscent of the 
anisotropy of the pseudogap observed experimentally in the 
insulating\cite{ronning} and in the superconducting\cite{timusk} cuprates,
but is not large enough to be comparable with that seen experimentally.
We speculate that the anisotropy seen here may be due to a difference
in the number of states near the Fermi energy at these two points in
the zone.

The DCA self-energy spectra Figs.~\ref{pizero} \& \ref{pio2pio2} support 
the spectral evidence. At the $X$ point, the slope of the real part 
${\rm Re}\Sigma({\bf{k}},\omega)$ becomes positive below $\beta=2.6$, the 
temperature at which we observed the opening of a pseudogap.  This signals 
the appearance of two new solutions in the quasiparticle equation 
${\rm{Re}}(\omega-\epsilon_{{\bf{k}}}-\Sigma({\bf{k}},\omega))=0$ in 
addition to the 
strongly damped solution at $\omega=0$ which is also present in the 
noninteracting system.  These two new quasiparticle solutions for the 
same $\bf k$-vector indicate precursor effects of the onset of 
antiferromagnetic ordering which entails a doubling of the unit cell. 
They are referred as shadow states and are caused by antiferromagnetic spin 
fluctuations in the paramagnetic state.  At these temperatures, the 
imaginary part ${\rm Im}\Sigma({\bf{k}},\omega)$ displays a local minimum at 
$\omega=0$ indicating the breakdown of the Fermi liquid behavior. We 
note that a different conclusion was previously reached in a FLEX 
study\cite{deisz}, which found that ${\rm Im}\Sigma({\bf{k}},\omega)$ has a 
local minimum at $\omega=0$ which was not accompanied by an opening of 
a pseudogap.  Since the pseudogap is due to short-range spin correlations, 
we conclude that FLEX underestimates these correlations.

\begin{figure}[ht]
\leavevmode\centering\psfig{file=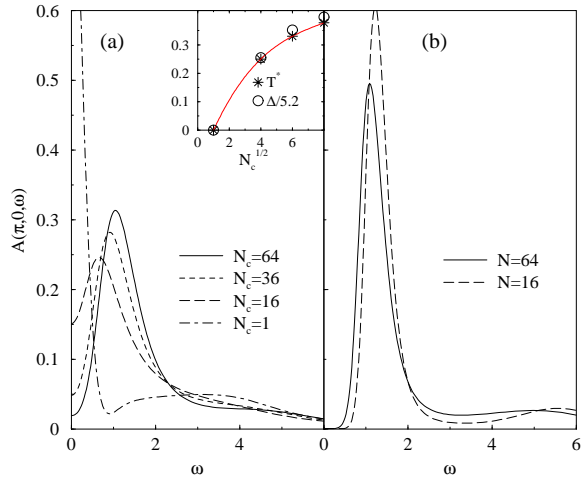,angle=-90,width=3. in}
\caption{The spectral density $A({\bf k}, \omega)$ at ${\bf k} = (\pi, 0)$ 
for the 2D Hubbard model via (a) the DCA and (b) finite-size Quantum Monte 
Carlo (QMC) at an inverse temperature times the bandwidth $\beta W = 40$ 
on various size clusters.  The temperature $T^*$ at which the pseudogap 
first becomes apparent in the DCA spectra, as well as the full width $\Delta$ 
measured from peak to peak is plotted in the inset.  The finite-size QMC
overestimates $\Delta$ and $T^*$, whereas the DCA QMC systematically 
underestimates them. 
}
\label{akwbss}       
\end{figure}

It is instructive to compare the DCA results with those obtained by 
finite-size QMC calculations.  Fig.~\ref{akwbss} shows the spectral 
density $A(\pi,0, \omega)$ obtained by analytically continuing 
both finite-size and DCA QMC data. In spite of the difference in the two 
methods, the information they provide is complimentary.  In the finite-size 
results, (b), we see a similar opening of a pseudogap.  However, as the 
length of the antiferromagnetic (AF) correlations reach the longest length on 
the finite-size lattice, the system develops a full gap.  Thus, the 
finite-size QMC overestimates the size of the gap.  In the DCA results, (a), 
the pseudogap emerges as soon as $N_c>1$.  The temperature $T^*$ at which 
the pseudogap first becomes apparent in the spectra, as well as the full 
width $\Delta$ measured from peak to peak is plotted in the inset.  
Both $T^*$ and $\Delta$ increase with $N_c$.  Since the DCA calculation
remains in the thermodynamic limit, a full gap due to antiferromagnetic 
correlations alone cannot open until their correlation length 
diverges.  However, since these correlations are restricted to the size 
of the cluster, the DCA systematically underestimates the size of the gap.
Thus, if a pseudogap exists in the DCA for finite $N_c$, it should persist 
in the limit as $N_c\to\infty$.  
	
In summary, we have employed the recently developed DCA to study the 
long-open question of whether the half-filled Hubbard model has a 
pseudogap due to AF spin fluctuations.  We find conclusive evidence of a 
pseudogap in the charge dynamics and have shown unambiguously that the 
$T=0$ phase transition of the half-filled model is preceded by an 
opening of a pseudogap in $A({\bf{k}}_F,\omega)$ accompanied by 
pronounced non-Fermi liquid behavior in $\Sigma({\bf{k}}_F,\omega)$.

\paragraph* {Acknowledgments}
We would like to acknowledge useful conversations with 
P.\ van Dongen,
B.\ Gyorffy,
M.\ Hettler,
H.R.\ Krishnamurthy
R.R.P.\ Singh
and
J.\ Zaanen.
This work was supported by the National Science Foundation grants
DMR-9704021, DMR-9357199, and the Ohio Supercomputing Center.

\end{document}